# Differential Evolution Algorithm Based Hyperparameter Selection of Gated Recurrent Unit for Electrical Load Forecasting

Anuvab Sen[1], Vedica Gupta[2], Chi Tang[3]

[1]Electronics Engineering Dept., IIEST Shibpur, India, [2]Economics Dept., Indian Institute of Technology Kharagpur, India,

[3]W Booth School of Engineering Practice and Technology, McMaster University, Canada.

Email address: cktang@mcmaster.ca

*Keywords—Load Forecasting, Differential Evolution, Gated Recurrent Unit, Meta-heuristics, Hyper-parameter Selection*

## Abstract

Accurate load forecasting plays a vital role in numerous sectors, but accurately capturing the complex dynamics of dynamic power systems remains a challenge for traditional statistical models. For these reasons, time-series models (ARIMA) and deep-learning models (ANN, LSTM, etc.) are commonly deployed and often experience higher success. In this paper, we analyze the efficacy of the recently developed Gated Recurrent Network model (GRU) in Load forecasting. GRU models have the potential to improve Load forecasting because of their ability to capture and model temporal dependencies in the data efficiently. We apply several metaheuristics namely Differential Evolution to find the optimal hyper-parameters of the Gated Recurrent Neural Network to produce accurate forecasts. Differential Evolution provides scalable, robust, global solutions to non-differentiable, multi-objective, or constrained optimization problems. Our work compares the proposed Gated Recurrent Network model integrated with different metaheuristic algorithms by their performance in Load forecasting based on numerical metrics such as Mean Squared Error (MSE) and Mean Absolute Percentage Error (MAPE). The simulation prediction is carried out by using the power load data of Ontario province, Canada. Our findings demonstrate the potential of metaheuristic-enhanced Gated Recurrent Network models in Load forecasting accuracy and provide optimal hyperparameters for each model.

## I. Introduction

Load forecasting is the application of science and technology to predict the future demand for electricity or power in a given geographical location, for some specific future time. It plays a crucial role in various sectors, such as energy trading & Markets, Infrastructure Planning, disaster management, etc., to name a few. Traditional load prediction methods rely on historical data & models that simulate patterns of electricity consumption, but such models often face challenges in accurately capturing the complex dynamics of power systems [1]. To model this complexity, time series models like Auto-Regressive Moving Average (ARIMA), various deep learning techniques have been introduced such as Artificial Neural Networks (ANN), Recurrent Neural Networks (RNN) and Long Short-Term Memory (LSTM). The models work to improve the accuracy of load forecasts by leveraging large datasets and discovering hidden patterns to predict future values. Like any other deep learning model, their performance depends on the chosen hyperparameters.

In this work we utilized metaheuristics Genetic Algorithm, Differential Evolution [2], and Particle Swarm Optimization to identify ideal hyperparameters. Although hyperparameter search techniques like Grid Search, Random Search, and Bayesian Optimization are substantial improvements to manual tuning, they are inferior to the metaheuristics discussed in this paper. The metaheuristics are more efficient than grid search and random search and more robust and scalable than Bayesian Optimization. Furthermore, these algorithms can be applied to nonlinear, nonconvex, and noncontinuous functions. In the field of load forecasting, metaheuristic algorithms were employed for the first time to optimize the performance of deep learning models like GRU. This resulted in the better performance of load forecasting models because metaheuristic algorithms excel in global optimization, exploring and exploiting the search space efficiently to find the optimal solution. In this paper, the authors seek to fill the void and propose Differential Evolution optimized Gated Recurrent Unit (GRU) architecture specifically designed for Load forecasting. We identify that the GRU's abilities in long-range dependencies can be applied to Load forecasting. This combination of GRU and Differential Evolution enables us to achieve enhanced accuracy and performance in predicting daily hourly demand. Additionally, we have collected and meticulously curated a comprehensive dataset comprising power-related parameters spanning a substantial time frame of 5 years specifically for the Ontario province in Canada.

The performance of the model was evaluated using Mean Squared Error (MSE) and Mean Absolute Percentage Error (MAPE) metrics. To evaluate the results, we also integrated Particle Swarm Optimization and Genetic Algorithm with the Gated Recurrent Networks to benchmark against our proposed Differential Evolution integrated Gated Recurrent Model.

Our work stands as the pioneering effort in proposing a Differential Evolution-based hyperparameter tuning scheme for Gated Recurrent Unit (GRU) Network models in the realm of short-term electrical load forecasting.

## II. CONTENTS

### A. Proposed Gated Recurrent Unit Model

The Gated Recurrent Network (GRU) [3] model's proposed structure is illustrated in Figure 1, highlighting its key components and connections. We have fed the model with 8 features from the previous three timesteps or from the data of the previous 3 hours including itself. So, we have used 36, 64 and 24 neurons for the input, GRU layer and output layers. For the case of GRU model architecture, we have taken the time steps into account, thus getting the shape or x-train: (53558, 3, 8) and likewise for the other splits.

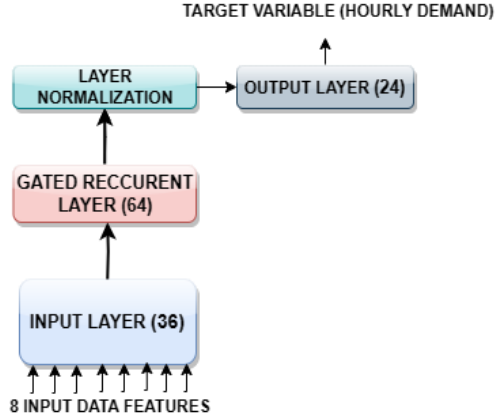

Fig.1. Proposed Gated Recurrent Model

All three networks used the Rectified Linear Unit (ReLU) activation function modeled in Equation 1 below:

$$f(x) = \max(0, x) \quad (1)$$

The metaheuristic algorithms were utilized to find the optimal batch size, epoch, and learning rate hyperparameters. They were optimized by minimizing loss using the Mean Squared Error metric. Each model's best set of learning rate, batch size and epoch are summarized in the table below:

| BEST SET OF HYPERPARAMETERS FOR GATED RECURRENT MODEL | |
|---|---|
| METAHEURISTICS | HYPERPARAMETERS |
| Genetic Algorithm | (80,844,0.0001) |
| Particle Swarm Optimisation | (173,35,0.3109) |
| Differential Evolution | (24,1000,0.1) |

Fig.2. Best set of hyper-parameters obtained for proposed Model

Mean Absolute Percentage Error (MAPE) was calculated for the best set of hyperparameters found using the Differential Evolution Algorithm.

### B. Differential Evolution Algorithm

Differential Evolution (DE) is a stochastic population-based optimization algorithm developed by Rainer Storn and Kenneth Price in 1997. It is used to find approximate solutions to a wide class of challenging objective functions. DE can be used on functions that are nondifferentiable, non-continuous, non-linear, noisy, flat, multi-dimensional, possess multiple local minima or are stochastic. The steps are explained below:

*1) Initialization:* Suppose f has D parameters. An N-sized candidate solution population is initialized, with each candidate solution modeled as $x_i$, a D-parameter vector.

$$x_{i,G} = [x_{1,i,G}, x_{2,i,G}...x_{D,i,G}], \text{ where } i = 1, 2...N, \text{ and } G \text{ is the generation number} \quad (2)$$

*2) Mutation:* A mutation is a stochastic change that expands the candidate solution search space. Mutations are used in DE to prevent the algorithm from converging upon a local optimum.

The mutant vector is obtained by adding the weighted difference of two of the vectors to the third.

$$v_{i,G+1} = v_{r1,G} + F \times (v_{r2,G} - v_{r3,G}) \quad (3)$$

$F \in [0, 2]$ represents the scale factor for the mutation.

*3) Crossover:* Crossover is how successful candidate solutions pass their characteristics to the following generations. A trial vector $u_{i,G+1}$ is created by combining the original vector $x_{i,G}$ and its corresponding mutant vector $v_{i,G+1}$. A widely used crossover scheme is described below:

$$u_{j,i,G+1} = \{v_{j,i,G+1}, \text{ if } p_{rand} U(0, 1) \leq CR, \text{ else: } x_{j,i,G}\} \quad (4)$$

*4) Selection:* Given both the initial target vector and generated trial vector, the fitness of each is evaluated using the initial objective or cost function f.

$$x_{i,G+1} = \{u_{i,G+1}, \text{ if } f(u_{i,G+1}) \leq f(x_{i,G}), \text{ else: } x_{i,G}\} \quad (5)$$

The Differential Evolution Algorithm is illustrated in Figure 3 below:

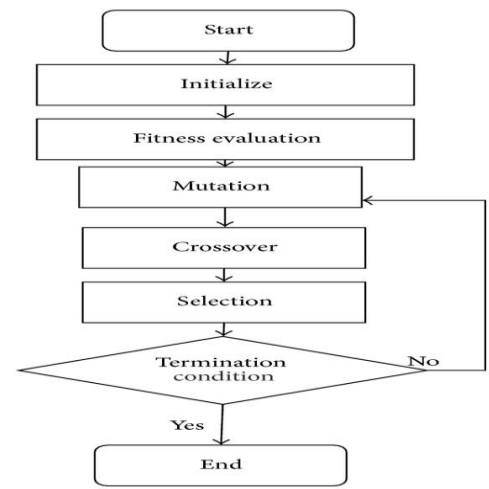

Fig.3. Differential Evolution Algorithm

Consequently, due to its power optimization capabilities, it was selected as the algorithm of choice to optimize our Gated Recurrent Unit (GRU) model.

*C. Data Collection*

The dataset was created for this project using data scraped from the official website of Government of Canada [4] and Independent Electricity System Operator (IESO), Canada [5]. This data contains meteorological data from Ottawa, Toronto, Ontario from 1st January 2017 to 31st July 2023. The data set contains the properties: Ontario hourly demand (in kW), Daily Peak Load (in kW), date, time, year, quarter, month, week of the year, day of the year, state holiday, an hour of the day, day of the week, day type, temperature (in ◦ Celsius), dew point temperature, relative humidity (in %), wind speed (in kilometers/hours), visibility (in kilometers) and precipitation amounts (in millimeters). The compiled data set contains 19 variables layered in 96,432 rows, where each row represents an hour.

## III. RESULTS AND DISCUSSION

We obtained the mean absolute percentage error (MAPE) using the proposed approach to implement the differential evolution-based hyperparameter tuning of the Gated Recurrent Network. This MAPE was compared to the MAPEs generated from the proposed approach with the genetic algorithm and particle swarm optimization-based hyperparameter tuning of the architecture.

The Standard scaler has been used to improve the convergence and stability of seasonal data during model training. The mean squared error (MSE) was employed as the loss function to evaluate the fitness of the differential evolution algorithm.

Mean Absolute Percentage Error (MAPE) is used to gauge the accuracy of the entire model. It provides a measure of the average percentage difference between predicted values and the actual values.

Table II above provides us with a comparison of MAPE among various metaheuristic optimization algorithms used here

| BEST SET OF HYPERPARAMETERS FOR GATED RECURRENT MODEL | |
|---|---|
| METAHEURISTICS | MAPE |
| Manual Selection | 2.07 |
| Genetic Algorithm | 1.31 |
| Particle Swarm Optimisation | 1.28 |
| Differential Evolution | 1.11 |

Fig.4. Comparision of MAPE & Meta-heuristics

The results prove that Differential Evolution (DE) algorithm outperforms the other meta-heuristics in terms of mean absolute percentage error (MAPE). Differential Evolution's superior ability can be attributed to a few factors. DE can more effectively explore the search space and exploit the promising regions for optimal solutions using its various genetic operators.

To visually understand the results and accuracy of the Load forecasting model proposed here we have plotted 24-hour prediction for the DE on the Gated Recurrent Network Model as shown in Figure 5. This shows that the model gives a fairly accurate prediction on Test data.

The results ascertain that metaheuristic optimization algorithms consistently outperform the manual selection method. Particle Swarm Optimization (PSO) outperforms Genetic Algorithm (GA) but falls behind Differential Evolution in overall optimization performance, showcasing DE's superior search capability and robustness.

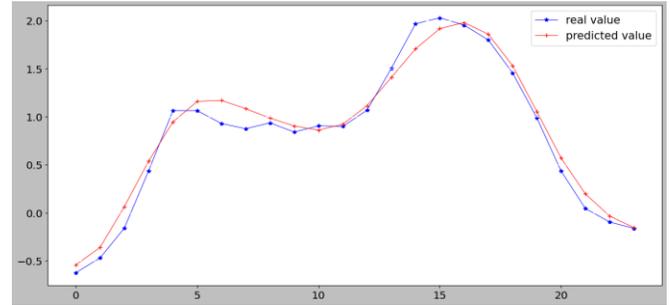

Fig.5. 24 hours ahead forecast plot for DE-GRU model

PSO suffers from rapid convergence, limiting its ability to reach the global optimum, which could be an explanation for its performance.

## IV. CONCLUSION

This paper applies several metaheuristic algorithms to a proposed Gated Recurrent Neural Network to find the optimal hyperparameters. This selection method was proven to be far more efficient and accurate than manual selection. Amongst the metaheuristics tested, Differential Evolution proved to be the best, followed by Particle Swarm Optimization and finally Genetic Algorithm.

Due to possessing limited computational resources, each metaheuristic algorithm couldn't be applied to sufficiently large populations over many generations. If this research is extended with more powerful devices, future studies over larger populations and more generations will corroborate our findings. Future study may investigate the performance of other alternative metaheuristic algorithms on hyperparameter tuning for similar deep learning models, across a wide range of forecasting tasks.

REFERENCES

[1] S. Sofi and I. Oseledets, "A Case Study of Spatiotemporal Forecasting Techniques for Load Forecasting," in IEEE Transactions on Power Systems, vol. 37, no. 4, pp. 1234-1256, 2022

[2] A. K. Das, S. M. Dash, and B. K. Panigrahi, "Differential Evolution: A Survey of the State-of-the-Art," in IEEE Transactions on Evolutionary Computation, vol. 15, no. 1, pp. 4-31, Feb. 2011

[3] Kyunghyun Cho, Bart van Merrienboer, Caglar Gulcehre, Dzmitry Bahdanau, Fethi Bougares, Holger Schwenk, and Yoshua Bengio. (2014). Learning Phrase Representations using RNN Encoder-Decoder for Statistical Machine Translation. In Proceedings of the Conference on Empirical Methods in Natural Language Processing (EMNLP), 1724-1734

[4] E.Canada and C. Change, "Government of Canada. Climate," Climate [Online]. Available: https://climate.weather.gc.ca/index_e.html

[5] IESO, "Demand Zonal Data," Independent Electricity System Operator. [Online]. Available: http://reports.ieso.ca/public/DemandZonal/

ACKNOWLEDGEMENT

The authors wish to express their gratitude to Mitacs Globalink & Dr. Chi Tang for sponsoring Anuvab Sen and Vedica Gupta to do undergraduate engineering research at McMaster University in the summer of 2023.